\documentclass[12pt]{article}
\usepackage[latin9]{inputenc}
\usepackage[a4paper]{geometry}
\usepackage{float}
\usepackage{amsmath}
\usepackage{amssymb}
\usepackage{graphicx}
\usepackage{setspace}
\usepackage[authoryear]{natbib}
\newtheorem{obs}{Observation}

\newtheorem{ass}{Assumption}
\newtheorem{res}{Result}

\usepackage{xcolor}
\definecolor{dark-red}{rgb}{0.4,0.15,0.15}
\definecolor{dark-blue}{rgb}{0.15,0.15,0.4}
\definecolor{medium-blue}{rgb}{0,0,0.5}

\usepackage{nicefrac}

\begin{document}

\title{An Ellsberg paradox for ambiguity aversion\thanks{We thank Michael Greinecker, John Nachbar, Paulo Natenzon, Andrea Prat, Todd Sarver, Tomasz Strzalecki, Peter Wakker, and Jonathan Weinstein for insightful comments. The authors gratefully acknowledge the funding from the Weidenbaum Center on the Economy, Government, and Public Policy at Washington University in St. Louis.}}

\author{Christoph Kuzmics\thanks{University of Graz, Austria, christoph.kuzmics@uni-graz.at} \and Brian W. Rogers\thanks{Washington University in St. Louis, U.S.A., brogers@wustl.edu} \and Xiannong Zhang\thanks{Corresponding author, Washington University in St. Louis, 1 Brookings Dr. St. Louis, Missouri, U.S.A. (63130), zhangxiannong@wustl.edu}}

\maketitle

\begin{abstract}
The 1961 Ellsberg paradox is typically seen as an empirical challenge to the subjective expected utility framework. Experiments based on Ellsberg's design have spawned a variety of new approaches, culminating in a new paradigm represented by, now classical, models of ambiguity aversion. We design and implement a decision-theoretic lab experiment that is extremely close to the original Ellsberg design and in which, empirically, subjects make choices very similar to those in the Ellsberg experiments. In our environment, however, these choices cannot be rationalized by any of the classical models of ambiguity aversion.
\end{abstract}


\newpage

\section{Introduction}

Ambiguity -- the kind of uncertainty that is not readily quantifiable with objective probabilities -- is ubiquitous in human decision-making and therefore of central importance to economic analysis. The classical approach to decision making under such uncertainty is via subjective expected utility, in which a decision maker (DM) makes their choices so as to maximize the expectation of utility relative to a subjective probability distribution. The experiments proposed by \citet{ellsberg61} have challenged this approach by finding observed behavior that is inconsistent with subjective expected utility.

The robust observation of such behavior in lab experiments has inspired more flexible models of decision-making that allow sensitivity to ambiguity and, in particular, which accommodate Ellsberg choices. Beyond the lab, recent work has sought to explain empirical phenomena with ambiguity-aversion, particularly in finance, see e.g., the surveys of \citet{epstein2010ambiguity} and \citet{guidolin2013ambiguity}.

In this paper we provide and execute an experimental design in which there exists an ``Ellsberg choice'', which, however, is \emph{incompatible} with any of the classical models of ambiguity aversion. With \emph{classical models of ambiguity aversion} we mean all those that satisfy a monotonicity axiom: If one act is better than another act in all states, then the inferior act cannot be chosen when the superior act is available.\footnote{\label{fn:ambiguitymodels}These include most preference models reviewed in the recent survey of \citet{machina2014ambiguity}, such as the maxmin expected utility model of \citet{gilboa89}, the Choquet expected utility model of \citet{schmeidler89}, the smooth ambiguity model of \citet{klibanoff05}, the variational and multiplier preference models of \citet{maccheroni06} and \citet{hansen2001robust}, confidence function preferences of \citet{chateauneuf2009ambiguity}, uncertainty aversion preferences of \citet{cerreia11}, and the incomplete preference model of \citet{bewley02}. There are several models of ambiguity aversion that do not insist on monotonicity, including \citet{seo09}, \citet{saito15}, \citet{bommier2017dual} and \citet{ke2020randomization}.}

Our main finding is that 66,67\% (18 out of 27) of our subjects make such Ellsberg choices,\footnote{The exact 95\% Clopper-Pearson confidence interval for the true proportion of such choices in the population is (46.04\%,83.48\%).} which is essentially the same as the frequency of Ellsberg behavior observed in traditional experiments.\footnote{Surveying 39 Ellsberg-type experiments, in addition to running their own, \citet{oechssler2015test} ``find that, on average, slightly more than half of subjects are classified as ambiguity averse.'' See also Table 3.4 in \citet{trautmann2015ambiguity}.  Studying uncertainty about correlations across random variables, \citet{epstein2019ambiguous} find similar empirical results regarding ambiguity aversion.} Thus, in very much the same sense that those experiments pose a challenge to the positive appeal of subjective expected utility, so does our experiment pose a challenge to the positive appeal of ambiguity aversion models.

Moreover, given the strong resemblance of our design to the traditional Ellsberg design, we argue that the most plausible explanation of Ellsberg choices is that they are driven by the same underlying forces in the two cases: it should be possible to understand the Ellsberg choice in our design and in the original design using the same (preference) model. This leads us to question the merit of existing ambiguity aversion models as the ideal explanation for Ellsberg behavior, since they cannot explain the fact that the same frequency of subjects make the Ellsberg choice in our experiment.

The paper proceeds as follows: The experimental design is presented in Section \ref{sec:design}. Section \ref{sec:theory} describes the necessary theoretical insights to interpret our findings. The results are then provided in Section \ref{sec:results}, with Section \ref{sec:disc} offering a brief discussion of these results. Section \ref{sec:lit} provides a more detailed discussion of the related literature, before Section \ref{sec:concl} offers concluding remarks. Experimental instructions and additional theoretical considerations are presented in the Appendix.

\section{Experimental Design} \label{sec:design}

The basis for our experiment is (a slight variation of -- to avoid indifferences) the \citet{ellsberg61} two-color urn experiment. There are two urns. The \emph{risky urn} contains 49 White and 51 Red balls. The \emph{ambiguous urn} contains 100 balls in total, each of which is either Green or Yellow; nothing more is known about the composition of the ambiguous urn. A decision maker (DM) is presented with a choice among three bets, which we shall call bets on White, Green and Yellow. The experimenter draws one ball from each of the two urns. Bet White pays out \$10 if the ball drawn from the risky urn is white and nothing otherwise. Bet Green (Yellow) pays out \$10 if the ball drawn from the ambiguous urn is green (yellow) and nothing otherwise. The Ellsberg choice is to Bet White. It is the only bet that cannot be accommodated by subjective expected utility.\footnote{If a subjective expected utility DM prefers White over Green, they must believe that drawing a Green ball is less likely than drawing a Yellow ball. If they also prefer White over Yellow, they must believe the opposite -- a contradiction.} Denote the set of bets by $\mathcal{B}=\{G,Y,W\}$.

In our experiment, before a subject makes their choice, they are shown two independent draws (with replacement) from the ambiguous urn. We refer to these as \emph{informational draws} because they do not directly affect payoffs. The possible informational draws are $\{GG,GY,YG,YY\}$, with $G$ for Green and $Y$ for Yellow. Each subject is asked to place a set of four conditional bets, i.e., to specify which bet (White, Green or Yellow) they would like to execute following each of the four possible informational draws. We call this profile of conditional bets a \emph{decision rule}. We elicit the decision rule before the informational draws are realized.\footnote{This is commonly referred to as the ``strategy method.''}

After the elicitation, the informational draws are executed and revealed to subjects. These determine the bet from the decision rule that is placed. We then draw one ball from each urn to determine the payment according to the chosen bet.\footnote{Complete experimental protocols and instructions are provided in the appendix. We confine the description in the main text to the basic procedures necessary to understand the experiments.}

\section{Theory} \label{sec:theory}

A representation of this experiment, in the spirit of \citet{anscombe63}, is as follows. The state space $\Omega=[0,1]$ is the set of all possible frequencies of green balls in the ambiguous urn. A state $\omega \in \Omega$ describes all non-objective uncertainty in the experiment. A decision rule is a function $\delta:\{GG,GY,YG,YY\}\to\mathcal{B}$ from the set of informational draws to the set of bets. We denote the decision rule that assigns $(GG\rightarrow a,GY\rightarrow b,YG\rightarrow c,YY\rightarrow d)$
by $(a,b,c,d)$ with $a,b,c,d\in\mathcal{B}$.

Every decision rule can be expressed as an act in the sense of \citet{anscombe63}, a function from $\Omega$ to objective probabilities of winning the prize. In state $\omega \in \Omega$, the probabilities of the four possible informational draws are $\omega^{2}$ for informational draws $GG$, $\omega(1-\omega)$ for informational draws $GY$, $(1-\omega)\omega$ for informational draws $YG$, and $(1-\omega)^{2}$ for informational draws $YY$. An act induced by decision rule $\delta=(a,b,c,d)$ attaches to each state $\omega\in\Omega$ the probability of receiving the price given by
\[
\pi(\delta|\omega)=\omega^{2}P(a|\omega)+\omega(1-\omega)P(b|\omega)+(1-\omega)\omega P(c|\omega)+(1-\omega)^{2}P(d|\omega),
\]
where $P(W|\omega)=49\%$, $P(G|\omega)=\omega$ and $P(Y|\omega)=1-\omega$.

A Bayesian (or subjective expected utility) DM holds a belief $\mu\in\Delta(\Omega)$, where $\Delta(\Omega)$ denotes the set of probability distributions over
$\Omega$, and chooses a decision rule according to $\max\mathbb{E}_{\mu}[\pi(\delta|\omega)]=\max\sum_{\omega\in\Omega}\pi(\delta|\omega)\mu(\omega)$.\footnote{As there are only two possible outcomes, maximizing the expectation of any utility function over outcomes is equivalent to maximizing the probability of the more preferred outcome.} The set of \emph{Bayesian decision rules} is, therefore, the set of rules for which there exists a belief $\mu\in\Delta(\Omega)$ under which the decision rule yields the highest expected probability of winning.\footnote{The following observations assume a non-degenerate prior. We rule out priors that attach probability one on state $\omega=0$ or probability one on state $\omega=1$. While we did not explicitly rule out such beliefs in our design, we feel that they are of limited practical relevance.}

\begin{obs} \label{obs:bayes} The set $\mathcal{S}$ of Bayesian decision rules
is given by
\[
\left\{ (G,G,G,G),(G,G,G,Y),(G,G,Y,Y),(G,Y,G,Y),(G,Y,Y,Y),(Y,Y,Y,Y)\right\} .
\]
\end{obs}

\noindent Proof: A Bayesian DM holds a prior $\mu_{0}\in\Delta(\Omega)$, which, after realization of the informational draws, is updated to some posterior belief $\mu_{1}\in\Delta(\Omega)$. But then, following exactly the same argument as in the canonical two-color Ellsberg environment, the DM must strictly prefer betting on either $G$ or $Y$ to betting on $W$ under belief $\mu_{1}$. No Bayesian rule can include a bet on $W$. Note next that Bayesian decision rules must be increasing in the number of $Y$'s in the informational draws: if a Bayesian DM bets on $Y$ for some informational draws they must also bet on $Y$ for any informational draws in which there are strictly more $Y$s. This follows from Bayesian updating. Finally, note that if the prior $\mu_{0}$ is symmetric around $\omega=1/2$ then the posterior beliefs following either $GY$ or $YG$ are also symmetric around $1/2$, and consequently $(G,G,Y,Y)$ and $(G,Y,G,Y)$ are both optimal decision rules. \hfill{$\square$}

We say that one decision rule $\delta$ \emph{dominates} another rule $\delta'$ if $\pi(\delta|\omega)\ge\pi(\delta'|\omega)$ for all $\omega\in\Omega$ with strict inequality for all $\omega\not\in\{0,1\}$.\footnote{While this is a slightly stronger form of dominance than weak dominance, we do not think the distinction is of any practical relevance in our context.} A decision rule $\delta$ is \emph{undominated} if there is no other rule $\delta'$ that dominates $\delta$. We say that an ambiguity aversion model is \emph{monotone} if it permits only undominated rules to be chosen, and refer to such models as \emph{classical}. We denote the set of dominated decision rules by $\mathcal{D}$ and the set of undominated decision rules by its complement $\mathcal{N}$.

\begin{obs} \label{obs:undom} Any decision rule that prescribes the bet $W$ after two (or more) of the four possible realizations of informational draws is dominated.
\end{obs}

\noindent Proof: To prove Observation \ref{obs:undom} we go through four cases.

Case 1: Consider a decision rule $\delta$ with $\delta(GG)=W$ and $\delta(GY)=W$ (the case of $\delta(YG)=W$ is analogous). Consider the decision rule $\delta'$ that is equal to $\delta$ except that $\delta'(GG)=G$ and $\delta'(GY)=Y$. Denote by $\Delta(\omega) = \pi(\delta'|\omega) - \pi(\delta|\omega)$ the expected difference in payoff between the two rules in state $\omega$. In the present case, $\Delta(\omega)= \omega^2 (\omega - \nicefrac{49}{100}) + \omega (1-\omega)((1-\omega) - \nicefrac{49}{100})$, or equivalently, $\Delta(\omega)= \omega \left(\omega^2 + (1-\omega)^2 - \nicefrac{49}{100} \left(\omega + (1-\omega) \right) \right)$. As $\omega + (1-\omega) = 1$ and $\omega^2 + (1-\omega)^2 \ge \nicefrac{1}{2}$ for all $\omega \in \Omega$ we have that $\Delta(\omega) \ge 0$ for all $\omega \in [0,1]$ and $\Delta(\omega) > 0$ for all $\omega \in (0,1]$. This proves that $\delta'$ dominates $\delta$.

Case 2: The case for decision rules $\delta$ such that $\delta(YY)=W$ and $\delta(YG)=W$ (or $\delta(GY)=W$) is analogous to the previous case. The rule $\delta$ is dominated by the rule $\delta'$ that is equal to $\delta$ except that $\delta'(YY)=Y$ and $\delta'(YG)=G$.

Case 3: Consider a decision rule $\delta$ with $\delta(GY)=\delta(YG)=W$. Consider the decision rule $\delta'$ that is equal to $\delta$ except that $\delta'(GY)=G$ and $\delta'(YG)=Y$. Then, $\Delta(\omega) = \omega (1-\omega) \left(1 - \nicefrac{98}{100} \right)$. Again we have that $\Delta(\omega) \ge 0$ for all $\omega \in [0,1]$ and $\Delta(\omega) > 0$ for all $\omega \in (0,1)$, and, therefore, $\delta'$ dominates $\delta$.

Case 4: Consider a decision rule $\delta$ with $\delta(GG)=\delta(YY)=W$. Consider decision rule $\delta'$ which is equal to $\delta$ except that $\delta'(GG)=G$ and $\delta'(YY)=Y$. Then, $\Delta(\omega) = \omega^3 + (1-\omega)^3 - \nicefrac{49}{100} \left(\omega^2 + (1-\omega)^2 \right)$. As $\omega^3 + (1-\omega)^3 \ge \nicefrac{1}{2} \left(\omega^2 + (1-\omega)^2\right)$ for all $\omega \in [0,1]$, we have that $\Delta(\omega) > 0$ for all $\omega \in [0,1]$ and, therefore, $\delta'$ dominates $\delta$. \hfill{$\square$}

In this decision problem, there are decision rules, such as $(G,G,G,W)$, that are undominated but not Bayesian. Every such rule prescribes the bet $W$ for exactly one informational draw. Most models of ambiguity aversion will not prescribe such behavior  for the given decision problem. A maxmin expected utility maximizer with a full set of priors, for instance, would choose the decision rule $(G,G,Y,Y)$ or $(G,Y,G,Y)$. These rules guarantee a winning probability of at least $\nicefrac{1}{2}$ in all states, while all rules involving a bet of $W$ yield a strictly lower worst case winning probability (in state $\omega = \nicefrac{1}{2}$) and all other Bayesian rules yield a lower worst case winning probability for some state $\omega \neq \nicefrac{1}{2}$.\footnote{Rule $(G,G,G,G)$ yields a winning probability of zero in state $\omega=0$. Analogously, rule $(Y,Y,Y,Y)$ yields a winning probability of zero in state $\omega=1$. Rules $(G,G,G,Y)$ and $(G,Y,Y,Y)$ yield a winning probability of $\nicefrac{13}{27} < \nicefrac{1}{2}$ in states $\omega=\nicefrac{1}{3}$ and $\omega=\nicefrac{2}{3}$, respectively.}

\section{Empirical findings} \label{sec:results}

The results of the experiment are summarized in Table \ref{tab:exp 2 resultb}.
Let us highlight the key finding.

\begin{res} \label{res:noambiguity} No decision rule in $\mathcal{D}$ is consistent with any of the classical models of ambiguity aversion (including subjective expected utility). Yet, the observed frequency of choices in $\mathcal{D}$ is $66.6\%$ (statistically significantly different from $0\%$). An exact Clopper-Pearson 95\% confidence interval for the true proportion of choices in $\mathcal{D}$ is (0.4604,0.8348).
\end{res}


\begin{table}[H]
\begin{centering}
\begin{tabular}{c|c|c}
Decision Rule  & Category  & Observation (percentage)\tabularnewline
\hline
GWWY  & $\mathcal{D}$  & 13(48.1\%) \tabularnewline
GGGY  & $\mathcal{S}$  & 5(18.5\%) \tabularnewline
WWWW  & $\mathcal{D}$  & 3(11.1\%) \tabularnewline
GGYY  & $\mathcal{S}$  & 2(7.4\%)\tabularnewline
WWWY  & $\mathcal{D}$  & 1(3.7\%) \tabularnewline
GGGG  & $\mathcal{S}$  & 1(3.7\%) \tabularnewline
GYWW  & $\mathcal{D}$  & 1(3.7\%) \tabularnewline
GYGY  & $\mathcal{S}$  & 1(3.7\%) \tabularnewline
\hline
\hline
 & $\mathcal{D}$  & 18(66.7\%)\tabularnewline
 & aWWd  & 17(63\%) \tabularnewline
Summary  & $\mathcal{S}$  & 9(33.3\%)\tabularnewline
 & aGYd(aYGd)  & 3(11.1\%)\tabularnewline
 & $\mathcal{N}\setminus\mathcal{S}$  & 0\tabularnewline
\end{tabular}
\caption{\label{tab:exp 2 resultb} Frequencies of decision rules from the experiment}
\end{centering}
\end{table}

\section{Discussion} \label{sec:disc}

\paragraph{Interpretation} The decision problem subjects face is close enough to that of the original Ellsberg two-color urn design, especially in the case of mixed informational draws (i.e., $GY$ or $YG$), that we argue that choices of $(aWWd)$ in our data, and choices of $W$ in the original Ellsberg design, are very likely to be governed by the same considerations. In fact, we have direct evidence to support this hypothesis. In a (non-incentivized) set of additional questions we asked subjects ``What choice would you have made if we had instead run the experiment without any informational draws?'' Of the 17 subjects who chose a rule of the form $(aWWd)$, 15 of them answered ``White.'' For comparison, of the 9 subjects who chose a decision rule compatible with subjective expected utility, only 2 answered ``White'', with the remaining 7 answering ``Green'', ``Yellow'', or ``Green or Yellow.''


\paragraph{State space} Most preference models concerning ambiguity are developed within the \citet{anscombe63} framework. This choice is most natural for our setting as well, as one can encode objective lotteries (through appropriate Anscombe-Aumann acts) and independent randomization (through the appropriate choice of a state being the urn composition rather than the color of the drawn balls). Our decision problem, can, however, also be modeled with a \citet{savage54} style state space and representation, in which states realize all uncertainty (including objective uncertainty). To do so, one needs to encode the extra information subjects are given through constraints on preferences.\footnote{\citet{lo00} has shown that no single choice can reveal violations of the \citet{savage54} axioms without restrictions on the subject's preferences.} Preferences need to reflect the subjects' knowledge of the risky urn composition and need to be exchangeable in the spirit of \citet{definetti1929funzione} (to capture the knowledge of independent draws from the urns), see also e.g., \citet{epstein2015exchangeable} and \citet{gilboa2016ambiguity}. A decision-maker with a preference that satisfies these two restrictions must strictly prefer any decision rule of the form $(aGYd)$ over the respective rule $(aWWd)$. The details of this argument are in the appendix. Thus, our conclusions are equally valid in the \citet{savage54} framework.\footnote{One could potentially take the position in the \citet{savage54} framework that incompatible behavior is the result of a subject's ``preferences'' not respecting their information about the risky urn's composition or the implications of exchangeability, rather than a failure of monotonicity.}


\paragraph{Timing} Our experiment has a sequential structure. First, there is a choice stage; second, informational draws are revealed; third, bets chosen in the first stage are executed; finally, the outcomes are realized and payments are made. Note, however, that there is only one choice stage. In contrast to genuinely dynamic environments with multiple choice stages, such a single choice-stage experiment presents no issues of a possible lack of commitment power or preference reversal and the (static) classical ambiguity aversion models apply directly. In the language of multiple selves models, it is here the time zero self that has to make decisions without any future selves making any decisions that could affect the welfare of the time zero self. It should, therefore, be exclusively the preferences of the time zero self that matter for their decision.\footnote{The literature on time-inconsistency typically assumes that earlier selves would like to restrict the choices of future selves, see e.g. \citet[Section 5.1]{frederick2002time}. Only in the absence of commitment possibilities may behavior exhibit a preference reversal. See \citet{siniscalchi2011dynamic} for consistent planning in dynamic decision problems under ambiguity, also \citet{epstein1993dynamically} and \citet{gilboa1993updating}, in generally dynamic settings.}



\paragraph{Complexity} One might argue that this experiment is significantly more complicated than the original Ellsberg experiment, and subjects' confusion in the former tells us little about the reasons for behavior in the latter. But few real-life decisions are as simple as the Ellsberg experiment and a useful theory should be widely applicable beyond such simple experiments. Moreover, the above mentioned evidence about the subjects' hypothetical choices without informational draws, indicates that our subjects understand the problem similarly to the original Ellsberg experiment.

\section{Related Literature} \label{sec:lit}


\citeauthor{machina09} (\citeyear{machina09,machina2014ambiguity2}) provides a number of thought experiments, which are then adapted and experimentally tested by \citet{schneider2018experimental}, in which classical ambiguity aversion models are unable to capture plausible behavior. These experiments are inspired by the \citet{allais53} example but adapted to a setting with subjective uncertainty and, thus, rely on a substantially richer environment than Ellsberg's examples, requiring, in particular, the use of at least three prizes. An important advantage of our design is that it remains as close as possible to Ellsberg's. This facilitates a comparison of our data to the existing body of data, and inherits some of the merits of Ellsberg's examples, including the robustness afforded by having only two prizes.

One of the closest papers to ours is \citet{halevy07}, who finds that the people who display ambiguity aversion are essentially the same people who fail to correctly reduce (objective) compound lotteries. As most models of ambiguity aversion are consistent with expected utility when there is only objective risk, the finding of \citet{halevy07} is evidence against such models. However, the association between ambiguity aversion and a failure to reduce compound lotteries is much weaker in the experiments of \citet{abdellaoui2015experiments}.

\citet{yang2017testing} introduce an experimental design based on the original Ellsberg two-color urn problem in which the risky urn has an equal number of balls of each color. Subjects can choose to bet on either the risky or the ambiguous urn. In either case, two balls will be drawn with replacement from the urn indicated by the subject. Then the DM is paid one prize $x$ if the first drawn ball is white (and nothing if red) and another prize $x$ if the second drawn ball is red (and nothing if white). Thus the DM can be paid $0,x$ or $2x$. Given this design both urns yield the same expected payoff with the risky urn yielding the highest variance. Thus, any risk averse DM, ambiguity averse or not, prefers the ambiguous urn. \citet{yang2017testing} find, however, that ``27-52\% {[}of{]} subjects in different treatments violated the predictions of subjective expected utility {[}and all classical ambiguity aversion models{]}.'' \citet{jabarian2022two} conduct a related experiment using the same urn setup, but where a subject must choose between the two urns and wins if two balls drawn with replacement are the same color.  The ambiguous urn dominates the risky urn, yet 55\% of their subjects bet on the risky urn.  Relative to these studies, an important advantage of our design is the close similarity to the original Ellsberg design. This allows us to demonstrate not only that there are settings in which classical ambiguity aversion models do not fare well, but to argue that these models are likely not the best explanation for Ellsberg choices even in the original experiments.

In a very different domain involving only risk, \citet{polisson2020revealed} make a point in a similar spirit as our work.  They show that the explanatory power of expected utility is nearly as good as a number of more general models among subjects who do not choose dominated lotteries.

\section{Conclusion} \label{sec:concl}

Most uncertainty that people face is difficult, if not impossible, to quantify objectively. Central to efforts to model, comprehend, and predict economic behavior is understanding the underlying principles that drive individuals' decisions in such environments. \citet{ellsberg61} identified robustly observed behavior that is inconsistent with the paradigm of subjective expected utility. As a response, more flexible preference models have been proposed that permit an aversion to ambiguity. These, by now classical, models of ambiguity aversion are consistent with the behavior Ellsberg has identified.

Our main conclusion is that these preference models face a similar descriptive problem, by virtue of our results, as subjective expected utility faces by virtue of the results from Ellsberg's original formulation.  Since the two decision-making environments are so closely related, it is unsatisfactory to require two distinct explanations, or preference models, to understand the data.  If one seeks a unified explanation of our data, alongside the classical results of Ellsberg choice experiments, that explanation  cannot be an aversion to ambiguity as represented in these classical models.


\label{sec:con}


\appendix

\section{Using a Savage style state space}

If we assume that subjects understand (and trust) the experimental instructions, certain information must be reflected in how we formally write down their decision problem. One way to do so is by using an Anscombe-Aumann state space representation, as we did in the main part of the paper. In particular this allowed us to capture two important facts: that balls are drawn independently from the urns, and that the risky urn contains 49 white and 51 red balls. These facts are encoded by taking states to be the urn composition and by an appropriate translation of decision rules to Anscombe-Aumann acts.

One could instead use a Savage style representation, in which states resolve all uncertainty, not only the subjective kind. For our problem this would mean that a state is a vector $s=(s_{1},s_{2},s_{3},s_{0})$, where, for $i=1,2,3$, $s_{i}\in\{\mathbf{G},\mathbf{Y}\}$ is the color of the $i$-th ball drawn from the ambiguous urn, and $s_{0}\in\{\mathbf{R},\mathbf{W}\}$ is the color of the ball drawn from the risky urn.\footnote{A state is therefore given by a quadruple of ball colors, such as $\mathbf{G}\mathbf{Y}\mathbf{G}\mathbf{R}$. We use this notation so as to not confuse states with decision rules, which are functions from the four possible realizations of the first two drawn balls from the ambiguous urn to bets in the set $\mathcal{B}=\{G,Y,W\}$ and which we, therefore, also depict by a quadruple such as GWWY, with the first entry the bet after the first two balls from the ambiguous urn realize in $\mathbf{G}\mathbf{G}$, the second after $\mathbf{G}\mathbf{Y}$, the third after $\mathbf{Y}\mathbf{G}$, and the fourth and last after $\mathbf{Y}\mathbf{Y}$.} Let $\Omega$ denote the set of all these states.

In our experiment subjects can only ever get a fixed monetary prize (call this $1$) or nothing (call this $0$). Each decision rule gives rise to a Savage act. A Savage act is a function $f:\Omega\to\{0,1\}$. Let $\mathcal{F}$ denote the set of all such acts and let $\succeq$
(with strict part $\succ$ and indifference part $\sim$) be a DM's (not necessarily complete, but transitive) preference relation over $\mathcal{F}$. While there are, thus, $2^{16}$ Savage acts, our subjects have to choose from the much smaller subset of acts that are induced by the $3^{4}$ feasible decision rules, which we denote by $\mathcal{F}^{*}.$

This state space representation, so far, does not encode all the information that the subjects were given. In particular it does not reflect the two vital pieces of information that balls are drawn independently from the urns, and that the risky urn contains 49 white and 51 red balls. This information will be encoded through the preference models that we allow decision makers to have.

As a background assumption (that we implicitly made throughout the paper), the DM prefers $1$ (the prize) over $0$ (not getting the prize). The following assumption captures what it means for the DM to believe that the ball drawn from the risky urn is more likely red than white.

\begin{ass} \label{ass:redmorelikely}
The preference $\succeq$ (with $\succ$ and $\sim$) \emph{reflects the risky urn information} if for any two acts $f,g\in\mathcal{F}$ such that there are $s_{1},s_{2},s_{3}\in\{\mathbf{G},\mathbf{Y}\}$ with $f(s_{1},s_{2},s_{3},\mathbf{R})=0$, $f(s_{1},s_{2},s_{3},\mathbf{W})=1$, $g(s_{1},s_{2},s_{3},\mathbf{R})=1$, $g(s_{1},s_{2},s_{3},\mathbf{W})=0$, and $f(s)=g(s)$ for all $s\in\Omega$ with $s\not\in\{(s_{1},s_{2},s_{3},\mathbf{R}),(s_{1},s_{2},s_{3},\mathbf{W})\}$, we have that $g\succ f$.
\end{ass}

Given a permutation (bijection) $\pi:\{1,2,3\}\to\{1,2,3\}$ and a (Savage) state $s=(s_{1},s_{2},s_{3},s_{0})\in\Omega$ let $s^{\pi}=(s_{\pi(1)},s_{\pi(2)},s_{\pi(3)},s_{0})$ denote the $\pi$-permutation of $s$. We call two acts $f,g\in\mathcal{F}$ \emph{exchangeable} if there is a permutation $\pi:\{1,2,3\}\to\{1,2,3\}$ such that $f(s)=g(s^{\pi})$ for all states $s\in\Omega$. The following assumption, appealing indirectly to de Finetti's theorem -- \citet{definetti1929funzione}, states what it means for the DM to understand that balls are drawn independently (from the ambiguous urn): the DM is indifferent between any two exchangeable acts.

\begin{ass} \label{ass:exchangeable}
The preference $\succeq$ (with $\succ$ and $\sim$) is \emph{exchangeable} if for any two exchangeable acts $f,g\in\mathcal{F}$ we have $f\sim g$.
\end{ass}

These two assumptions still do not reflect all the information that the DM is given. For instance, they do not reflect that the balls from one urn are drawn independently from those drawn from the other urn. But these assumptions suffice for our argument.

\begin{obs} \label{obs:savagedom}
For any (not necessarily complete, but transitive) preference relation $\succeq$ (with strict part $\succ$ and indifference part $\sim$) over the set of Savage acts $\mathcal{F}$ that reflects the risky urn information and is exchangeable, $(aGYd)\succ(aWWd)$ for any $a,d\in\mathcal{B}$.
\end{obs}

To see this consider the following table, in which we compare the two decision rules in those states for which the two decision rules provide different outcomes.

\begin{table}[H]
\begin{centering}
\begin{tabular}{c|cccc}
(Savage) state  & aWWd  & D  & aGYd  & \tabularnewline
\hline
$\mathbf{G}\mathbf{Y}\mathbf{G}\mathbf{W}$  & 1  & 1  & 1  & \tabularnewline
$\mathbf{G}\mathbf{Y}\mathbf{G}\mathbf{R}$  & 0  & 0  & 1  & \tabularnewline
$\mathbf{G}\mathbf{Y}\mathbf{Y}\mathbf{W}$  & 1  & 0  & 0  & \tabularnewline
$\mathbf{G}\mathbf{Y}\mathbf{Y}\mathbf{R}$  & 0  & 1  & 0  & \tabularnewline
$\mathbf{Y}\mathbf{G}\mathbf{G}\mathbf{W}$  & 1  & 0  & 0  & \tabularnewline
$\mathbf{Y}\mathbf{G}\mathbf{G}\mathbf{R}$  & 0  & 1  & 0  & \tabularnewline
$\mathbf{Y}\mathbf{G}\mathbf{Y}\mathbf{W}$  & 1  & 1  & 1  & \tabularnewline
$\mathbf{Y}\mathbf{G}\mathbf{Y}\mathbf{R}$  & 0  & 0  & 1  & \tabularnewline
\end{tabular}
\par\end{centering}
\centering{}\caption{\label{tab:savagestatespacechoice} Comparison of decision rules $(aGYd)$
and $(aWWd)$ in Savage state space $\Omega$.}
\end{table}

The act induced by decision rule $(aGYd)$ can be obtained from that induced by decision rule $(aWWd)$ in two steps. First we produce decision rule $D$ from $(aWWd)$ by swapping the outcomes in $(aWWd)$ in states $\mathbf{G}\mathbf{Y}\mathbf{Y}\mathbf{R}$ and $\mathbf{G}\mathbf{Y}\mathbf{Y}\mathbf{W}$ and in states $\mathbf{Y}\mathbf{G}\mathbf{G}\mathbf{R}$ and $\mathbf{Y}\mathbf{G}\mathbf{G}\mathbf{W}$. By Assumption \ref{ass:redmorelikely} the DM strictly prefers $D$ over $(aWWd)$. Then we produce decision rule $(aGYd)$ from $D$ by swapping the outcomes in $D$ in states $\mathbf{G}\mathbf{Y}\mathbf{Y}\mathbf{R}$ and $\mathbf{Y}\mathbf{G}\mathbf{Y}\mathbf{R}$ and in states $\mathbf{Y}\mathbf{G}\mathbf{G}\mathbf{R}$ and $\mathbf{G}\mathbf{Y}\mathbf{G}\mathbf{R}$. By Assumption \ref{ass:exchangeable} the DM is indifferent between decision rules $D$ and $(aGYd)$. Together with transitivity this implies that the DM strictly prefers $(aGYd)$ over $(aWWd)$ if the DM has exchangeable preferences and considers red more likely than white.

Observation \ref{obs:savagedom}, thus, implies that any subject with a preference $\succeq$ that reflects the risky urn information and is exchangeable, when faced with the set of Savage acts $\mathcal{F}^{*}$ induced by the available decision rules in the experiment, cannot choose any decision rule of the form $(aWWd)$ as it is not in the set of $\succeq$-maximal acts within $\mathcal{F}^{*}$.

This implies that the same 63\% of subjects' choices of the form $(aWWd)$, see Table \ref{tab:exp 2 resultb}, can also not be explained by any preference in the Savage framework that reflects the risky urn information (Assumption \ref{ass:redmorelikely}) and is exchangeable (Assumption \ref{ass:exchangeable}). Put differently, any preference model that makes a choice of the form $(aWWd)$ optimal in the given decision problem, must violate either exchangeability or transitivity, or ignores the risky urn information.

Note that these two assumptions are compatible with ambiguity aversion. Ellsberg behavior, such as choosing $W$ in the baseline treatment of our second experiment, is not a violation of Savage preferences that satisfy these assumptions. To see this, note that we can embed our slightly modified two-color Ellsberg treatment in the given Savage state-space representation as follows. Consider three decision rules, labelled $W,G$ and $Y$, summarized in the following table.

\begin{table}[H]
\begin{centering}
\begin{tabular}{c|cccc}
(Savage) state  & W  & G  & Y  & \tabularnewline
\hline
$\mathbf{Y}\mathbf{G}\mathbf{G}\mathbf{W}$  & 1  & 1  & 0  & \tabularnewline
$\mathbf{Y}\mathbf{G}\mathbf{G}\mathbf{R}$  & 0  & 1  & 0  & \tabularnewline
$\mathbf{Y}\mathbf{G}\mathbf{Y}\mathbf{W}$  & 1  & 0  & 1  & \tabularnewline
$\mathbf{Y}\mathbf{G}\mathbf{Y}\mathbf{R}$  & 0  & 0  & 1  & \tabularnewline
\mbox{%
all other states%
}  & 0  & 0  & 0  & \tabularnewline
\end{tabular}
\par\end{centering}
\centering{}\caption{\label{tab:savagestatespacechoice2} Assumptions \ref{ass:redmorelikely}
and \ref{ass:exchangeable} do not rule out ambiguity aversion.}
\end{table}

Assumptions \ref{ass:redmorelikely} and \ref{ass:exchangeable} have no bite in this example. One cannot obtain one of the three rules from another by a combination of permuting the first three coordinates of the state and shifting the prize from states with a $\mathbf{W}$ in the fourth coordinate to states that differ only by having $\mathbf{R}$ in the fourth coordinate. Thus, there is a preference $\succeq$ that reflects the risky urn information and is exchangeable that satisfies $W\succ G,Y$. On the other hand, for the usual reasons, no subjective expected utility DM could choose $W$ when choosing from $W$, $G$, and $Y$. For more on modelling exchangeable preference with ambiguity aversion see \citet{epstein2015exchangeable}.

\section{Experimental Design}

\subsection{Experiment details}

The experimental sessions took place in April and May of 2018. The experiment was conducted at Missouri Social Science Experimental Laboratory (MISSEL) at Washington University in St. Louis. 31 students, 4 of them acting as monitor, participated in the experiments and the average session length was 45 minutes.

\begin{figure}[H]
\centering{}\includegraphics[scale=0.3]{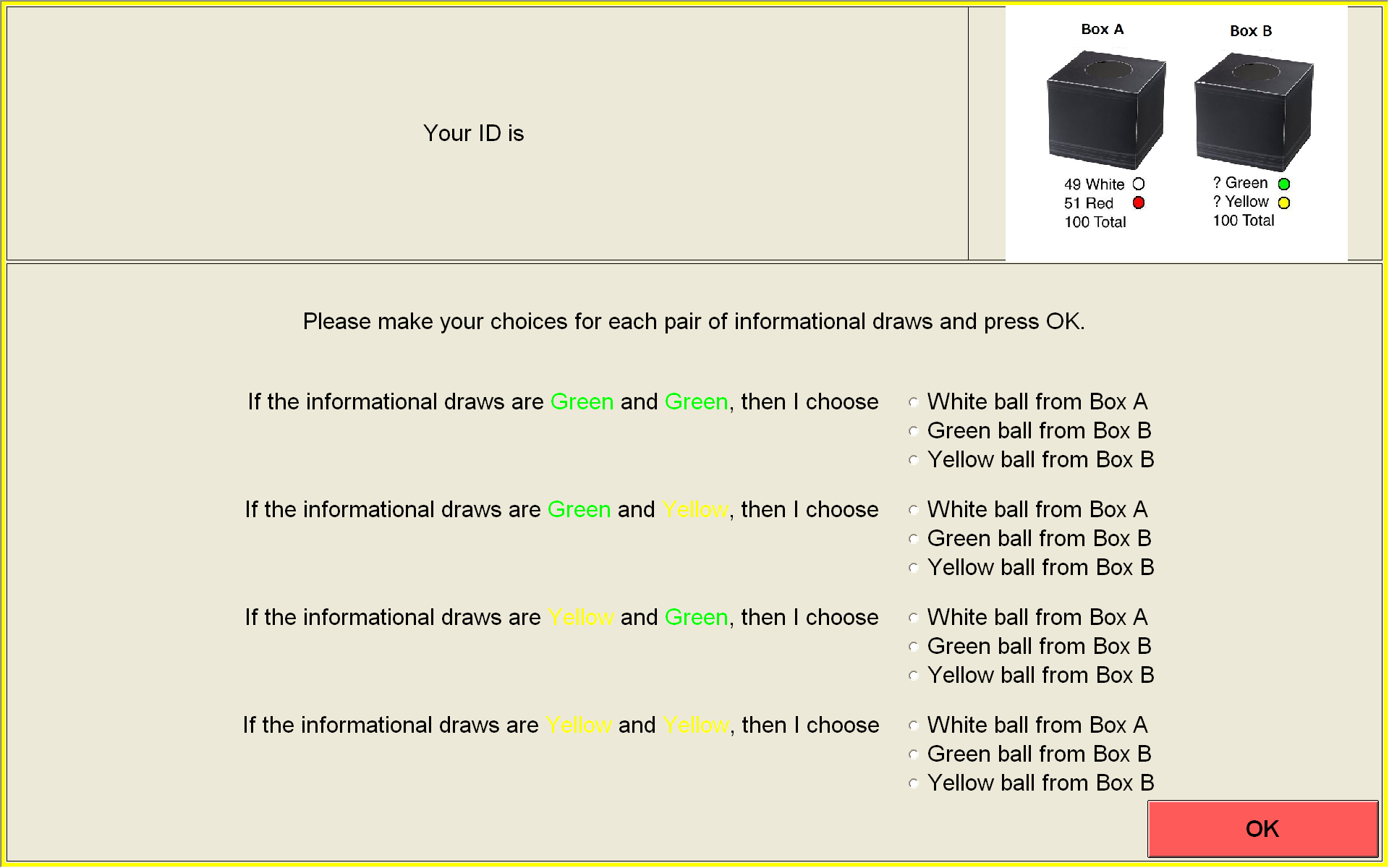}\caption{\label{fig:SS3} Strategy method}
\end{figure}

\begin{figure}[H]
\centering{}\includegraphics[scale=0.3]{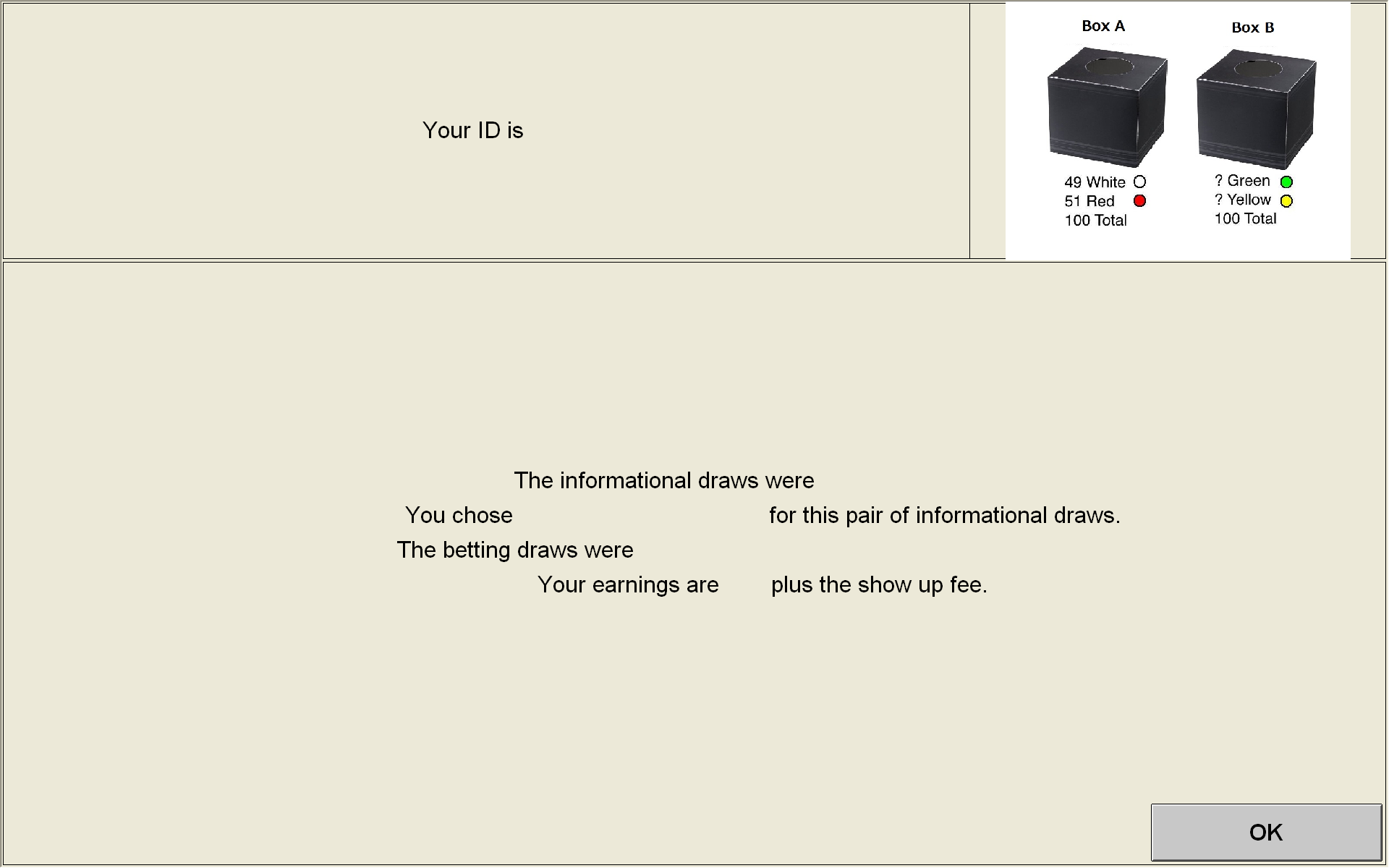}\caption{\label{fig:SS4} Payoff screen}
\end{figure}

The subjects answered exactly one incentivized question, which was related to guessing the color of a ball. If the guess was correct, the subject received 10 USD, and 0 otherwise. The show-up fee was 5 dollars. At the end of the session we conducted a short questionnaire. The questions were not incentivized, but we emphasized that answering these questions would be helpful for our research. The experiment was programmed using z-Tree \citep{fischbacher2007z}. See Figures \ref{fig:SS3} and \ref{fig:SS4} for screen shots.

\subsection{Physical environment}

\begin{figure}[H]
\centering{}%
\begin{tabular}{cc}
\includegraphics[scale=0.045]{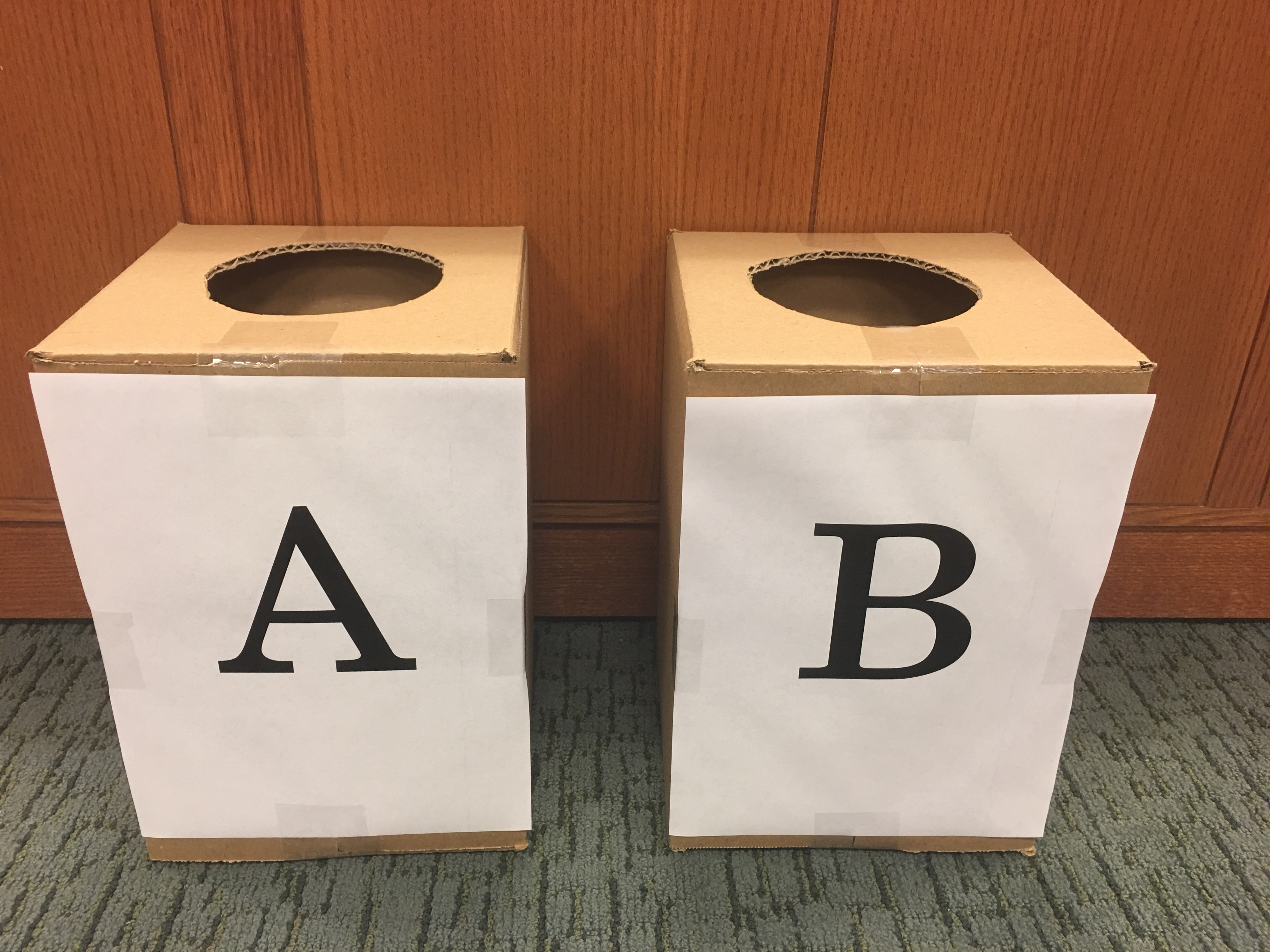}  & \includegraphics[scale=0.045]{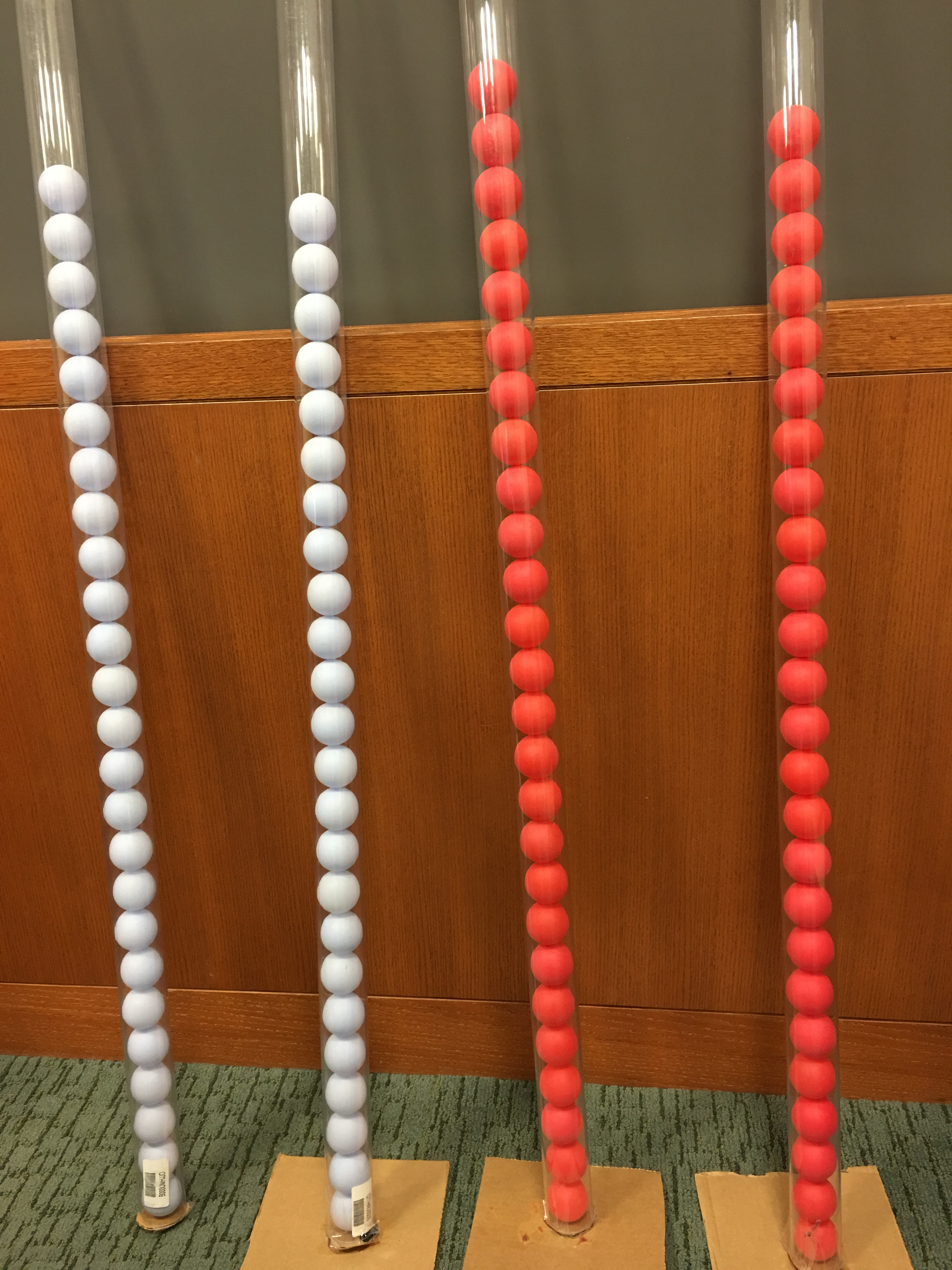}\tabularnewline
a  & b\tabularnewline
\end{tabular}\caption{\label{fig:Boxes}Boxes}
\end{figure}

The urns and states were implemented using two cardboard boxes and colored ping-pong balls. During the experiment (and in what follows), we refer to the two containers as Box A and Box B. A photo of the boxes can be found in Figure \ref{fig:Boxes} (a). The protocols we used were guided by the desire to be as clear and transparent as possible. Box A contained 49 white and 51 red balls. The balls were displayed in clear plastic tubes at the beginning of the experiment so that subjects could easily see that there were two more red than white balls. Photos of the tubes are included as Figure \ref{fig:Boxes} (b). After showing the balls to subjects, they were poured into Box A. On the other hand, it was important that the exact contents of Box B were unknown. We therefore informed subjects that Box B contained 100 balls, each of which was either green or yellow, but we were intentionally not telling them anything further about the contents. Box B was shaken so that it was credible that it contained the same number of balls as Box A. After this presentation, subjects were told that they could inspect all the boxes and balls at the conclusion of the experiment if they so desired.

In each session, one subject was randomly selected to act as a monitor. The monitor was the person who physically conducted all draws of balls and displayed their colors to the other subjects.

\subsection{Screen Shots}

\subsection{Questionnaire}

Table \ref{tab:Questionnaire-3} shows the questions that we asked at the end of the experiment.

\begin{table}[H]
\centering{}%
\begin{tabular}{|c|}
\hline
Gender (Male, Female, Prefer not to tell) \tabularnewline
Major \tabularnewline
How many Green balls do you think there are in Box B? \tabularnewline
How many Yellow balls do you think there are in Box B? \tabularnewline
What choice would you have made if we had instead run the experiment \tabularnewline
without any informational draws? \tabularnewline
Please tell us how you made your choice for informational draws \tabularnewline
``Green \& Yellow'' and ``Yellow \& Green.'' \tabularnewline
Please tell us how you made your choice for informational draws \tabularnewline
``Green \& Green'' and ``Yellow \& Yellow.'' \tabularnewline
Was there any part of the experiment that was unclear? \tabularnewline
\hline
\end{tabular}\caption{List of questions asked in the questionnaire of the first experiment\label{tab:Questionnaire-3}}
\end{table}


\subsection{Instructions}

\begin{center}
\textbf{\large{}{}{}{}{}{}{}{}{}{}{}{}{}Instructions}{\large\par}
\par\end{center}

\begin{center}

\par\end{center}

Welcome to the experiment! Please take a seat as directed. Please wait for instructions and do not touch the computer until you are instructed to do so. Please put away and silence all personal belongings, especially your phone. We need your full attention for the entire experiment. Adjust your chair so that you can see the screen in the front of the room. The experiment you will be participating in today is an experiment in decision making. At the end of the experiment you will be paid for your participation in cash. Each of you may earn different amounts. The amount you earn depends on your decisions and on chance. You will be using the computer for the experiment, and all decisions will be made through the computer. DO NOT socialize or talk during the experiment. All instructions and descriptions that you will be given in this experiment are factually accurate. According to the policy of this lab, at no point will we attempt to deceive you in any way. Your payment today will include a \$5 show up fee. One of you will be randomly selected to act as a monitor. The monitor will be paid a fixed amount for the experiment. The monitor will assist us in running the experiment and verifying the procedures. If you have any questions about the description of the experiment, raise your hand and your question will be answered out loud so everyone can hear. We will not answer any questions about how you ``should'' make your choices. As I said before, do not use the computer until you are asked to do so. When it is time to use the computer, please follow the instructions precisely.

We will now explain the experiment. There are two containers on the table that we will refer to as Box A and Box B. This is Box A. The Box is empty. Box A will contain 100 ping pong balls. Each of the balls in Box A will be either White, like this, or Red, like this. Specifically, Box A will contain exactly 49 White balls and 51 Red balls, for a total of 100 balls. You don't have to remember these numbers. When it is time to make a decision, we will remind you of these numbers. We have counted and displayed the balls in these tubes to make it easier to show the contents of Box A. There are 25 white balls in this tube and 24 in this tube, for a total of 49 white balls. There are 25 red balls in this tube and 26 red balls in this tube, for a total of 51. We will now pour these balls into Box A and shake it to mix the balls together. This is Box B. We have already filled Box B with 100 ping pong balls. Each ball is either Green, like this, or Yellow, like this. We will not reveal the exact numbers of Green and Yellow balls. Instead, you know only that there are 100 balls in total, consisting of some combination of Green and Yellow balls. We will now shake Box B to mix the balls up. At the end of the experiment, you will have an opportunity to inspect the Boxes and ping pong balls if you wish. In a few moments we will ask the Monitor to draw one ball from each Box for everyone to observe. You will be asked to choose from several options that correspond to guessing the color of a ball that the Monitor draws. If your guess matches the result, you will receive 10 dollars in additional to the show up fee. If your guess does not match, you will receive 0 dollars in addition to the show up fee.

We will now start the experiment. On the computer desktop you will find a green icon named zleaf. Double click it now. Now there should be a welcome screen. Type your name and click the OK button in the welcome screen. One of you has been randomly selected by the software to serve as the monitor. Please raise your hand if you are the monitor. Could you please click the OK button on your screen and come to the
front? Now your screen should have changed to ``Please listen to the instructions.'' Please leave it like that and do not click OK. In a few moments the Monitor is going to draw one ball from Box A and one ball from Box B. We are going to ask you to bet on the outcome of those draws.

Now, recall that Box A contains 49 white balls out of 100 balls. Box B contains an unknown combination of Green and Yellow balls. Before we ask you to place your bet, we are going to show you two draws from Box B. We call these the ``informational draws'', since they are just for your information. To do this, the Monitor will first draw a ball and show it to everyone. We will then return the ball, shake Box B and have the monitor draw and display a second ball. The second ball will then be returned to Box B so that it still has the same 100 balls. So, these informational draws have four possible outcomes: Green and Green, Green and Yellow, Yellow and Green, Yellow and Yellow. In a few minutes you will see a screen like this. We are going to ask you to consider which bet you want to choose in each of those four cases. Even though only one of your bets will be implemented, you need to answer all four of these questions.

For example, the first question asks: ``If the information draws are Green and Green, which bet do you want to choose?'' What this means is the following. Suppose that when the Monitor conducts the informational draws from Box B, s/he draws first a Green Ball, the replaces it, and then draws another Green ball. Now you have to choose a bet. Which bet do you want to make if the informational draws turn out to be Green and Green? The other questions are similar, and ask you to choose which bet you want to place for the other possible informational draws. The second question asks you which bet you want to place if the informational draws are first Green and then Yellow. The third question asks you which bet you want to place if the informational draws are first Yellow and then Green. The fourth question asks you
which bet you want to place if the informational draws are first Yellow and then Yellow. After you have made all four choices, we will then have the monitor actually conduct the informational draws. Once the informational draws are revealed, this will determine which one of your four choices will be implemented. So, even though you answered all four questions, we will implement only your bet that corresponds to the actual informational draws. Once your bet has been determined according to the informational draws, the monitor will then conduct the betting draws by drawing a single ball from Box A and a single ball Box B to determine if your bet wins or loses. To summarize, you choose a bet for all four possible cases, then the monitor will conduct the two informational draws from Box B. This determines your bet. Finally, the monitor will conduct the betting draws by drawing one ball from Box A and one ball from Box B. The betting draws will determine whether your bet wins or loses. Please make your choices on the computer now. Then click the OK button and wait for others.

The monitor is now going to draw the balls. We will now conduct the informational draws. Please look away and draw a ball from Box B and show it to everyone. The color is {[}REALIZED COLOR{]}. Please put the ball back. We will write down the result on the blackboard. Now please look away and draw a second ball from Box B and show it to everyone. The color is XXX. Please put the ball back. We will write down the result on the blackboard. The informational draws are {[}REALIZED COLOR{]} and {[}REALIZED COLOR{]}. So the bet we will implement is your choice corresponding to information draws {[}REALIZED COLOR{]} and {[}REALIZED COLOR{]}. Now it is time for betting draws. Please look away and draw a ball from Box A and show it to everyone. The color is {[}REALIZED COLOR{]}. Please put the ball back. We will write down the result on the blackboard. Now please look away and draw a ball from Box B and show it to everyone. The color is {[}REALIZED COLOR{]}. Please put the ball back. We will write down the result on the blackboard. Now please return to your seat and enter these results into your computer screen, accompanied by an Experimenter. You can now see the outcome and your earnings on the screen. If you have questions about your payoff, please raise your hand.

We will now conduct a short questionnaire. Please wait for the questionnaire to start. The monitor doesn't have to fill the questionnaire. Please complete the questionnaire. Please be as specific as you can in your responses. Answering the question is helpful to our research, but your responses are entirely voluntary. After you finish, please wait for others. We will call you to the front by your participant ID to be paid before leaving. Thank you very much for your participation. This concludes the experiment. We will now begin calling you to the front to be paid before leaving.

 \bibliographystyle{abbrvnat}
\bibliography{reference}

\end{document}